\documentclass[sigconf]{acmart}

\AtBeginDocument{%
  }

\setcopyright{acmlicensed}

\copyrightyear{2025}
\acmYear{2025}
\setcopyright{acmlicensed}\acmConference[ICCPS '25]{ACM/IEEE 16th
International Conference on Cyber-Physical Systems (with CPS-IoT Week 2025)}{May 6--9, 2025}{Irvine, CA, USA}
\acmBooktitle{ACM/IEEE 16th International Conference on Cyber-Physical Systems (with CPS-IoT Week 2025) (ICCPS '25), May 6--9, 2025, Irvine, CA, USA}
\acmDOI{10.1145/3716550.3722019}
\acmISBN{979-8-4007-1498-6/2025/05}

\usepackage{titlesec} 

\usepackage{hyperref}
\usepackage{subcaption}
\usepackage{multirow}
\usepackage{enumitem}
\usepackage{amsmath}
\usepackage{xcolor}

\begin{document}

\title{ Psychophysiology-aided Perceptually Fluent Speech Analysis of Children Who Stutter}


\author{Yi Xiao}
\affiliation{%
  \institution{Arizona State University}
  \city{Tempe}
  \country{USA}}
\email{yxiao124@asu.edu}

\author{Harshit Sharma}
\affiliation{%
  \institution{Arizona State University}
  \city{Tempe}
  \country{USA}}
\email{hsharm62@asu.edu}

\author{Victoria Tumanova}
\affiliation{%
  \institution{Syracuse University}
  \city{Syracuse}
  \country{USA}}
\email{vtumanov@syr.edu }

\author{Asif Salekin*}
\affiliation{%
  \institution{Arizona State University}
  \city{Tempe}
  \country{USA}}
\email{Asif.Salekin@asu.edu}
\thanks{\textbf{(*)} Corresponding author.}
\thanks{\textbf{Acknowledgements:} This work was partly supported by NIH NIDCD \# R21DC018103, NSF IIS SCH \#2124285 and NIH R01 \#R01DC020959.}


\begin{abstract}
This paper presents a novel approach named \textit{PASAD} that detects changes in perceptually fluent speech acoustics of young children. Particularly, analysis of perceptually fluent speech enables identifying the speech-motor-control factors that are considered as the underlying cause of stuttering disfluencies. Recent studies indicate that the speech production of young children, especially those who stutter, may get adversely affected by situational physiological arousal. A major contribution of this paper is leveraging the speaker’s situational physiological responses in real-time to analyze the speech signal effectively. The presented \textit{PASAD} approach adapts a \textit{Hyper-Network} structure to extract temporal speech importance information leveraging physiological parameters. 
Moreover, we collected speech and physiological sensing data from $73$ preschool-age children who stutter (\textit{CWS}) and who do not stutter (\textit{CWNS}) in different conditions. \textit{PASAD}’s unique architecture enables identifying speech attributes distinct to a \textit{CWS}’s fluent speech and mapping them to the speaker’s respective speech-motor-control factors. Extracted knowledge can enhance understanding of children’s speech-motor-control and stuttering development. Our comprehensive evaluation shows that \textit{PASAD} outperforms state-of-the-art multi-modal baseline approaches in different conditions, is expressive and adaptive to the speaker’s speech and physiology, generalizable, robust, and is real-time executable.
\end{abstract}

\ccsdesc[500]{Computing methodologies~Neural networks}
\ccsdesc[500]{Applied computing~Health informatics}
\ccsdesc[500]{Human-centered computing~Ubiquitous and mobile computing systems and tools}

\keywords{Stuttering, Children Who Stutter, Machine learning, Explainable AI}

\maketitle

\section{Introduction}
Approximately $5$ percent of all children go through a period of stuttering, and $1$ percent suffer from long-term stuttering \cite{stutterpopulation}.
Stuttering first emerges in preschool years and, for those children who develop chronic stuttering, it can lead to life-long adverse consequences and compromised wellbeing \cite{carter2017self}. Research \cite{heyde2016fluent,usler2017lag} suggests that among other factors, the audible speech disfluencies in stuttering individuals result from underlying differences in \textit{speech-motor-control} processes \cite{amster1987articulatory} that are present in the fluent speech of people who stutter.
Importantly, preschool-age is when children rapidly develop their language, speech-motor-control, and emotion regulation. Studies \cite{yaruss2002national} have established that early-age interventions are highly effective in permanent remissions of stuttering \cite{curlee1993early,curlee1993evaluating}. Therefore, there is a pressing need for automated assessment tools capable of detecting subtle differences in the \textit{perceptually fluent speech} of children who stutter (\textit{CWS}). Such tools could enable just-in-time, personalized interventions at an early age. This study represents a first step in that direction.

 However, audio-based models alone struggle to differentiate between typical developmental disfluencies and persistent stuttering due to variability in speech patterns and situational influences \cite{jones2014temperament}. Speech science research has demonstrated that situational physiological responses impact speech-motor control in individuals who stutter \cite{erdemir2018effect,diehl2019situational}. Consequently, integrating psychophysiology-inspired inputs linked to speech-motor control offers complementary information, enhancing classification accuracy—particularly in ambiguous cases where perceptually fluent speech may still reflect underlying speech-motor difficulties \cite{jones2014autonomic,smith2017stuttering}.

\textit{Building on this insight, a key contribution of this work is the real-time incorporation of a speaker’s situational physiological arousal to analyze speech signals. This approach significantly improves the accuracy of classifying perceptually fluent speech as originating from either \textit{CWS} or children who do not stutter (\textit{CWNS}).} Notably, the real-time inference of the classifier enables effective just-in-time interventions.  
\par 
Conventional sequential classifiers (e.g., LSTM, GRU) used in speech analysis literature \cite{shewalkar2019performance} use an identical set of network weights in each timestamp. However, children's speech gets affected by their psychophysiology \cite{erdemir2018effect}, meaning each timestamp’s acoustic analysis should be adaptive to the speaker’s physiological response on that timestamp, that the conventional classifiers’ static network weights fail to facilitate. 
The paper presents a \emph{first-of-its-kind} Psychophysiology aided Speech Attribute Detection (\textit{PASAD}) approach that differentiates perceptually fluent speech from \textit{CWS} vs. \textit{CWNS}. \textit{PASAD} addresses the above-discussed challenge by presenting a unique \emph{\textit{HyperNetwork} \cite{ha2016hypernetworks} adaption} that leverages the speaker’s real-time psychophysiological parameters to update a sequential speech classifier’s network weights dynamically. Hence, the classifier has a new set of weights at each timestamp, meaning it can focus on (i.e., analyze) different aspects of speech acoustic conforming to the speaker’s real-time psychophysiology.
\par 
\textit{PASAD} processes speech input using a Mel-spectrogram representation. Since Mel-spectrogram features are distributed across various time and frequency ranges, capturing relationships between distant features necessitates a deep network architecture. However, this poses a challenge in speech health applications, where training data is often limited. To overcome this limitation, \textit{PASAD} incorporates a non-local block, enabling it to effectively capture long-range spatiotemporal relationships in the Mel-spectrogram while making efficient use of limited training data.

\par 
Later, utilizing machine learning (ML) interpretation techniques, we visualize the \textit{distinct speech attributes} of \textit{CWS} that are indicative of their speech-motor-control differences from the \textit{CWNS}. Conventional sensor data integration approaches concatenate the multi-modal sensing data within their network architectures and generate inferences utilizing the integrated information. Hence, it is difficult to identify the independent contribution of one modality from such architectures utilizing ML interpretation techniques.
\emph{\textit{PASAD}’s unique architecture} utilizes only speech information for inference generation, enabling using ML interpretation techniques like Kernel SHAP \cite{lundberg2017unified} to extract the distinct speech attributes present in \textit{CWS}'s speech and their corresponding speech-motor-control factors. 
\par 
The main contributions and empirical findings of this work are:
\begin{itemize}[noitemsep,nolistsep]
    \item \textbf{Classification:} \textit{PASAD} uses a \textit{Hyper-LSTM} unit that dynamically prioritizes the speech attributes taking the speaker’s real-time psychophysiology into account.
    Our comprehensive evaluations demonstrate that \textit{PASAD} is expressive and adaptive to the speaker’s speech and physiological parameter changes, utilizes only speech information in the class inference generation, is generalizable to different unseen age and sex groups, and is real-time executable in resource constraint devices.
    \item \textbf{Datasets:} To our knowledge, no existing public dataset contains both speech and physiological parameters of \textit{CWS} and \textit{CWNS}. We collected \emph{two first-of-their-kind datasets, `\textit{stress-speech} dataset’ and `\textit{narrative-speech} dataset’} comprising \textit{CWS} and \textit{CWNS}’s perceptually fluent speaking and physiological parameters in stressful condition and spontaneous narration task, respectively. The narration task is linguistically and cognitively demanding since the children develop new contexts and articulate them in speech. It may elicit different physiological responses in \textit{CWS} vs. \textit{CWNS} \cite{zengin2018sympathetic}. Two datasets comprise pre-school age (i.e., $3$-$5$ years) $73$ participants ($34$ \textit{CWS} and $39$ \textit{CWNS}) and would be made public with the published paper. 
    \item \textbf{Visualization:} \textit{PASAD}’s unique architecture enables visualization of distinct speech attributes present in \textit{CWS}’s speech and their corresponding speech-motor-control factors. Such visualization would enable remote and real-time assessment of \textit{CWS}’s speech-motor-control and would lead to near-real-time personalized, just-in-time interventions. \textit{PASAD} provides deeper insights into the underlying mechanisms of stuttering and has the potential to transform the early intervention for stuttering.
    
\end{itemize}

\section{Related Work and Background Discussion}\label{related-work}
\paragraph{\textbf{\textit{Studies in Speech Science:}}}
Physiological arousal-related differences in speech-motor-control have long been observed in the adults who stutter (\textit{AWS}) and don't stutter (\textit{AWNS}) \cite{van2014impact,jackson2016impact}. 
Limited research examined the effects of physiological arousal on speech characteristics of young children. According to \citet{tumanova2020effects}, \textit{CWS} demonstrate less stable speech-motor-control than \textit{CWNS} in emotionally arousing states. Despite this emerging evidence, the effects of emotional processes on perceptually fluent speech in \textit{CWS} and \textit{CWNS} are still not well understood.   
\par
\textbf{\textit{Machine Learning on Stuttering Speech Studies:}}
\label{ML-studies}
State-of-the-art ML studies on the stuttering population focus on detecting various stuttering disfluency events, such as repetition and prolongation of sounds/syllables, tension-filled pauses, and interjections. \citet{ravikumar2009approach} evaluated MFCC speech features and SVM classifier to detect syllable repetitions. Recent studies are leveraging deep learning approaches. 
\citet{kourkounakis2020detecting} developed a residual network (ResNet) and Bidirectional LSTM model, which took spectrogram as input to detect stuttering events. \citet{sheikh2021stutternet} utilized a time-delay neural network (TDNN), taking MFCC features to detect stuttering events. The majority of the stuttering speech processing works have used the UCLASS dataset \cite{howell2009university}. It contains stuttering disfluency event speech from participants aged $5-47$ years. In contrast, the data used for this paper are perceptually fluent speech and physiological parameters from preschool-age ($3$-$5$ years) \textit{CWS} and \textit{CWNS}. 
To our knowledge, no study has developed ML classification approaches to differentiate perceptually fluent speech of \textit{CWS} vs. \textit{CWNS}.
\par
\textbf{\textit{Speech Features in Stuttering and Arousal Studies:}}
Speech science studies have evaluated the fundamental frequency $F0$ and the first four formant frequencies ($F1$-$F4$) to examine the speech-motor-control both in emotional stimuli scenarios and for stuttering individuals.
\textit{$F0$} approximates the periodicity of vibration of the vocal cord \cite{guo2011design}. 
\textit{AWS} and \textit{AWNS} show difference in $F0$ while fluently uttering vowels \cite{salihovic2009characteristics}. 
\textit{Formants} are resonance frequencies of the vocal tract. They are indicative of time-dependent positions of speech articulators, e.g.,  tongue, lips, and jaw \cite{ladefoged1978generating}. 
Studies have shown that stress stimuli cause a rise in the first formant $F1$ and a fall in the second formant $F2$ values and narrow formant bandwidths \cite{banse1996acoustic,protopapas1997fundamental}. Additionally, the difference in the $F1$ is observed in \textit{AWS} vs. \textit{AWNS}s’ speech in stress conditions \cite{caruso1994adults}. $F2$ has been used to assess motor control of stuttering adults \cite{robb1998formant} and children \cite{bauerly2019effects}. \citet{scherer2013facets} have found significant differences in the $F2$ values for vowels and third formant $F3$ frequency fluctuations \cite{giannakakis2019review} between the neutral and aroused utterances. 
Additionally, Mel-frequency cepstral coefficient (MFCC) and zero crossing rate (ZCR) features have been leveraged in speech emotion detection \cite{salekin2017distant}. 
\par
\textbf{\textit{Physiological Parameters in \textit{CWS} vs. \textit{CWNS}:}}
Recent studies \cite{zengin2018sympathetic,tumanova2020emotional,sharma2022a} have shown significant skin conductance (EDA), heart rate (HR), and respiratory effort (RSA) differences in \textit{CWS} and \textit{CWNS} in different stress-inducing conditions. 

\section{Dataset Description and Collection Procedure}\label{DATASET}
We collected data from preschool-age children in two settings. Due to the nature of the tasks the participants were engaged in, we refer to the first setting dataset as the ``\textit{stress-speech} dataset'' and the second dataset as the ``\textit{narrative-speech} dataset'' throughout the manuscript. Data collection with this very young population can be difficult as they may refuse to do tasks. Presented clinical data collection took multiple years for our multi-disciplinary team.
\par
\textbf{Participants Description :}
Participants were between 3-5 years of age. 
The X University Institutional Review Board approved the study. Informed consent was obtained from parents, and verbal assent was obtained from children. Participants' speech and language skills were assessed using standardized speech articulation and language measures. Participants' speech fluency was assessed to diagnose developmental stuttering using evidence-based diagnostic criteria \cite{yaruss1998evaluating}. All participants had English as their primary language, age-appropriate speech articulation, and language scores, and passed a pure-tone hearing screening.
\par 
\textit{\textit{Stress-speech Dataset}} has $38$ participants. $18$ children ($16$ males and $2$ females, mean age $4$ years $5$ months) were \textit{CWS}, and $20$ were \textit{CWNS} ($17$ males and $3$ females, mean age $4$ years $6$ months).
\textit{\textit{Narrative-speech} Dataset} has $35$ participants. $16$ children ($13$ males and $3$ females, mean age $3$ years $11$ months) were \textit{CWS}. Other $19$ were \textit{CWNS} ($12$ males and $7$ females, mean age $4$ years $1$ month).
\par
\textbf{\textit{Extracted Physiological Sensors and Parameters: }}The respiratory, electrodermal, and cardiac activity were acquired simultaneously using the \textit{Biopac MP150 hardware system} \cite{biopac} and acqknowledge software (ver. 4.3 for PC, Biopac). Hypoallergenic electrodes were attached to the skin of the distal phalanges of the participants’ index and middle finger of the left hand for acquisition of electrodermal activity (EDA) and to the skin at the suprasternal notch of the rib cage and at the 12th rib laterally to the left for acquisition of the electrocardiogram (ECG) \cite{venables1980electrodermal}. A strain gauge transducer designed to measure respiratory-induced changes in thoracic or abdominal circumference (model TSD201, Biopac Systems, Inc.) was used to record respiratory effort (RSP). The transducer was positioned around the participants’s chest. All signals were sampled at $1250$ Hz. We extract the respiratory rate (RSP-rate), and respiratory amplitude (RSP-amp) \cite{schneider2003respiration} from raw RSP signal, and heart rate (HR) from raw ECG signal. 
Hence, the raw physiological parameters extracted and utilized are EDA, HR, RSP-rate, and RSP-amp.
\par
\textbf{\textit{Extracted Speech Signal:}} In each of the data collection setups, a \textit{Shure sm58 microphone} \cite{shure} was positioned approximately 50 cm (1.64 feet) from the participant’s mouth to record the participant’s speech. The speech was collected with a 10kHZ sampling rate.

\subsection{Sensor Data Collection Procedure}\label{Dataset-procedure}
Both datasets' data collection comprised a baseline and an experimental condition; the baseline was the same in both.
\par
\textbf{\textit{Baseline Condition (Common): }}
For both datasets, to establish a pre-experimental baseline for each participant’s resting psychophysiological parameters, participants viewed an animated screensaver of a three-dimensional fish tank for four minutes. This procedure has been successfully implemented in prior studies to establish baseline psychophysiological levels in preschool-age children \cite{jones2014autonomic}. 
\par
\textbf{\textit{\textit{Stress-speech} Experimental Condition: }} 
This data collection session lasted approximately $10$ minutes. Participants were shown $10$ negatively-valenced pictures from the International Affective Picture System \cite{lang1997international} and were asked to repeat a simple phrase ``Buy Bobby a puppy'' (BBAP) three times after each picture shown.
Participants (both \textit{CWS} \& \textit{CWNS}) did not show any stuttering events during this task, and their speech was entirely fluent.
\par
\textbf{\textit{\textit{Narrative-speech} Experimental Condition: }} The session lasted approximately 12 minutes and comprised a picture description task. Participants were shown pictures from a wordless storybook (\textit{Frog Goes to Dinner} by \citet{mayer1974frog}). To keep the narrative elicitation procedure consistent between participants, the examiner was only allowed to prompt the participant by saying, ``Let's look at this picture. Tell me what is happening here.’’ \textit{CWS} participants did show some stuttering events during this task, but they were infrequent. On avg., 97\% of their words were produced fluently. This study evaluated only the fluent speech portions.

\section{Features Extraction}\label{Features}
In the stress-speech dataset, each BBAP utterance takes 3-5 sec. Since, during SMC assessment, the existing speech science studies evaluate complete utterances, to be consistent, this paper detects \textit{CWS} vs. \textit{CWNS} speech from $5$ sec windows. From the \textit{stress-speech} and \textit{narrative-speech} datasets, a total of $1177$ ($581$ \textit{CWS} and $596$ \textit{CWNS}), and $3771$ ($1648$ \textit{CWS} and $2123$ \textit{CWNS}) non-overlapping $5$ sec windows were extracted. Each of the $5$ sec speech and physiological signals was divided into $19$ segments with $500$ ms duration and $250$ ms overlap, and physiological and acoustic features were extracted from the $500$ ms segments.
\vspace{-0.3cm}
\subsection{Physiological Features}
\label{physio-features}
Two categories of physiological features: (a) raw features and (b) changed-score features are evaluated. The presented best approach (i.e., \textit{PASAD}) uses change-score features as the input physiological feature representation.
\par
\textbf{\textit{Raw physiological features:}} Following previous studies \cite{salekin2017distant,partila2015pattern}, we consider the low-level descriptor (LLD) features: HR, EDA, RSP-rate, and RSP-amp from each $500$ ms segments. $6$ high-level descriptors (HLD) functionals: min, max, std, var, mean, and median are applied on the LLDs to extract the feature representation of each $500$ ms segment. We extract $19\times24$ raw physiological features from the 5-sec window.
\par
\textbf{\textit{Change-score Features:}} Following studies on \textit{CWS}'s physiological response \cite{sharma2022a,jones2017executive}, we extract change-scores of HR, EDA, RSP-amp, and RSP-rate LLD features from each $500$ ms segment. Change-scores are the differences between the $500$ ms physiological signal segments in different speech conditions from the same individual’s avg. physiological signal in the baseline (i.e., resting) condition. In this study, the differences (i.e., change-scores) are measured by two matrices: cosine similarity and the euclidean distance. For each of the four LLD features HR, EDA, RSP-amp, RSP-rate, we calculate two difference measures, totaling eight change-score features extracted from each $500$ ms segment. We extract $19\times8$ change-score features from the $5$ sec window.
\par
An individual may have an above-average physiological response (e.g., HR, EDA) in the baseline condition. However, a classifier trained on raw physiological features would be unaware of such exceptions and infer the individual's neutral state as a stress state. The change-score features eliminate such biases. A detailed discussion of change-score feature extraction is in Appendix \ref{appendix:feature}.

\vspace{-0.3cm}
\subsection{Acoustic Features}\label{acustic-features}
The presented approach extracts a Mel-spectrogram from each $500$ ms segment and detects $5$ sec speech categories (\textit{CWS} vs. \textit{CWNS}) from the sequence of $19$ spectrograms. To compare with speech literature, we also extracted and evaluated raw acoustic features: the first 13- Mel-frequency cepstral coefficient (MFCC), Zero crossing rate, fundamental frequency, and first four formant frequencies. A detailed discussion of the raw acoustic features is in Appendix \ref{appendix:feature}.
\par
\textbf{Mel-Spectrogram:} It represents the acoustic signal on the Mel scale. The Mel Scale is a logarithmic transformation of a signal’s frequency, such that sounds within an equal distance on the Mel Scale are perceived as equal to humans. 
\par 
\begin{figure}[h]
\centering
\includegraphics[width=7cm]{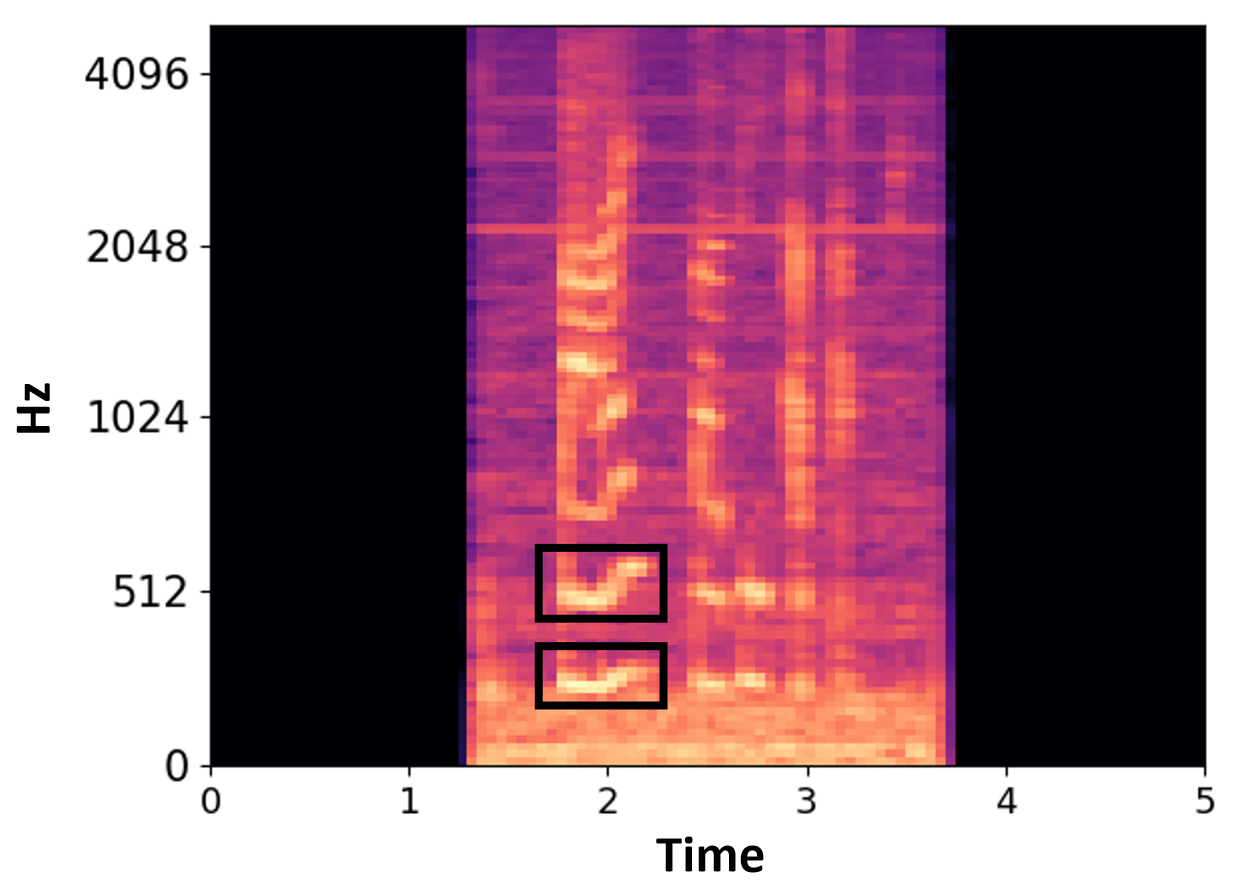}
\vskip -1ex
\caption{Mel-spectrogram for a $5$ sec speaking window}
\label{fig:spec}
\vskip -4ex
\end{figure}
The \textit{Fig.} \ref{fig:spec} shows a Mel-spectrogram of a `Buy Bobby a puppy’ \textit{stress-speech} utterance from a participant. Time and frequency are along the horizontal and vertical axis. Amplitude is represented by the \textbf{dark-to-bright} color, where the brighter pixels represent higher amplitude regions.
Mel-spectrogram contains information about the acoustic features studied by speech pathologists. For example, the $250$-$400$Hz region represents the children's speech fundamental frequency $F0$ information. Formants are seen as the \textbf{bright bands}. They are the peak points of the amplitude in the frequency spectrum.
The first two formants are highlighted in the bounding boxes in \textit{Fig.} \ref{fig:spec}.  The formants and their relations represent speech signal characteristics. Further detailed discussion is in Appendix \ref{appendix:feature}.

\section{Methodology: Psychophysiology-aided Speech Attribute Detection (\textit{PASAD})}

\begin{figure}[ht]
\vskip -2ex
\centering
\begin{subfigure}{0.49\textwidth}
  \centering
  \includegraphics[width=0.9\linewidth]{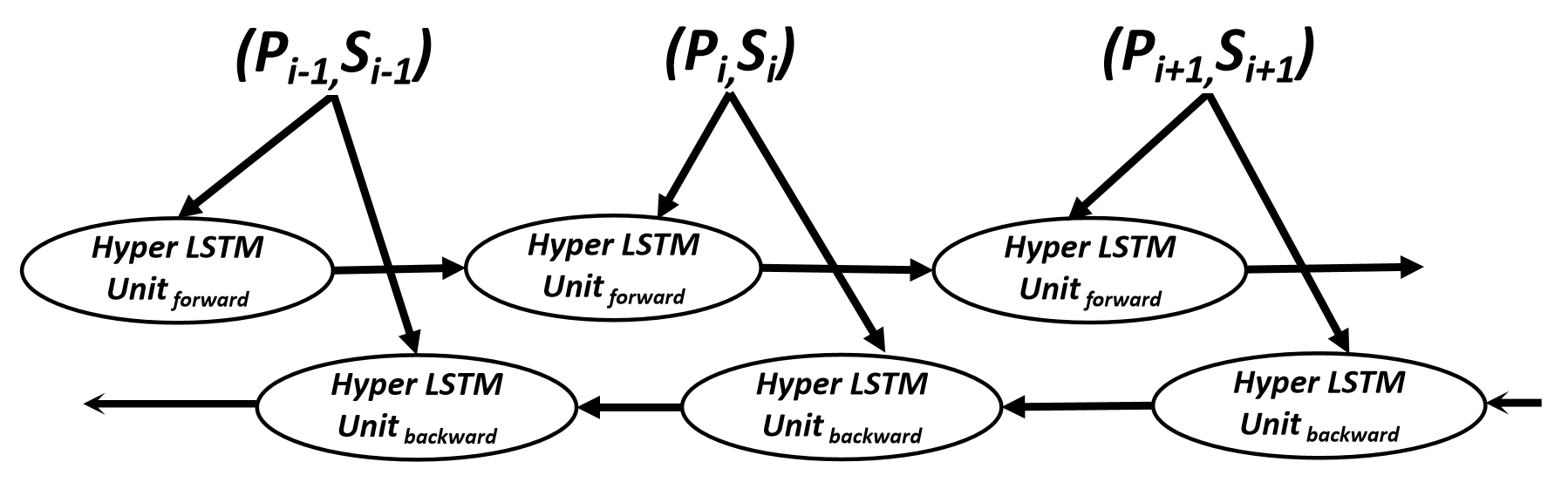}
  \vskip -1ex
  \caption{Overall Architecture.}
 \label{fig:bilstm_a}
\end{subfigure}
\begin{subfigure}{.49\textwidth}
  \centering
  \includegraphics[width=.9\linewidth]{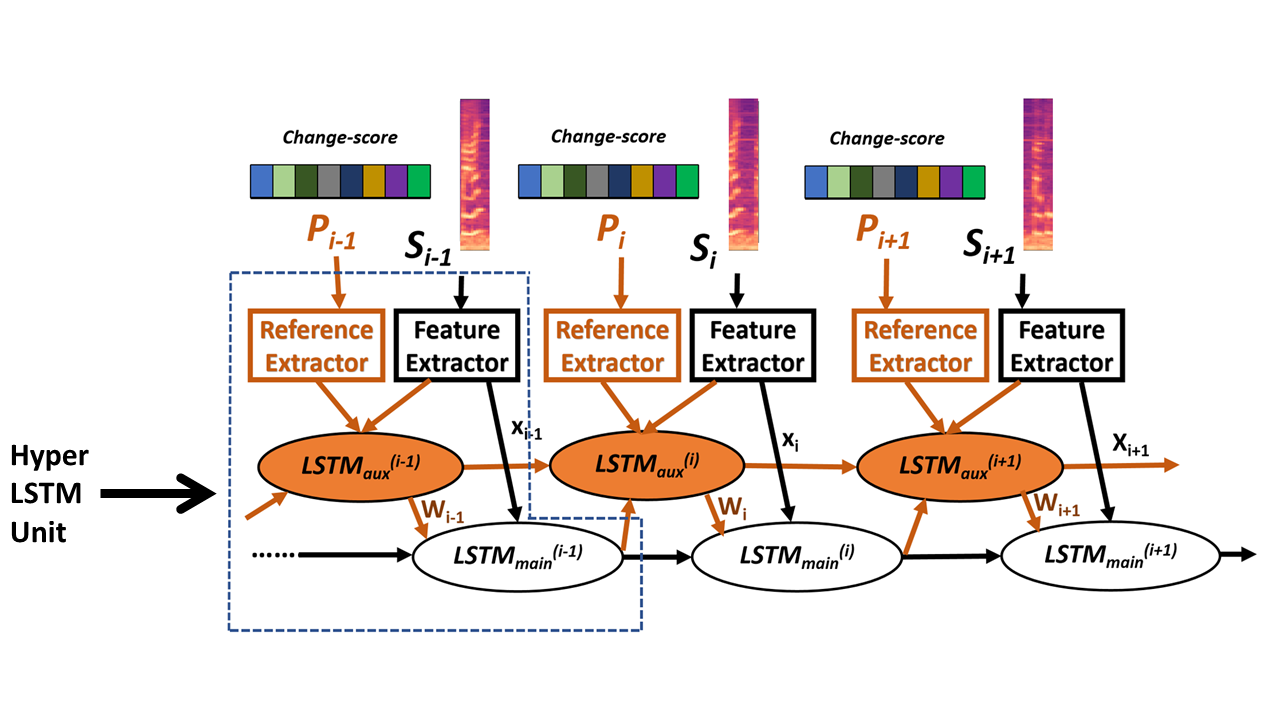}
  \caption{\textit{Hyper LSTM} unit.}
  \label{fig:bilstm_b}
\end{subfigure}
\vskip -2ex
\caption{\textit{PASAD}'s architecture.}
\label{fig:test}
\vskip -3ex
\end{figure}

\textbf{\textit{Overview:}}
\textit{PASAD} utilizes a novel \textit{Hyper-LSTM} unit (shown in \textit{Fig.} \ref{fig:bilstm_b}) that includes four components: reference-extractor, feature-extractor, auxiliary $LSTM_{aux}$, and main $LSTM_{main}$. At each timestamp $t$, the feature-extractor and the reference-extractor block generate the embeddings of the $500$ ms Mel-spectrogram and psychophysiological parameters. Classification inferences generating main-$LSTM_{main}$ take only the Mel-spectrogram embedding as input.
\par
Conventional LSTM unit \cite{cheng2016long} comprises four gates. 
Combined, these gates identify the importance of the input $x_t$ at timestamp $t$ compared to the overall sequential data.
These gate weights are fixed across timestamps during evaluation. Hence, LSTM is incapable of dynamic input importance extraction.
\par
In the presented \textit{Hyper-LSTM} unit (\textit{Fig.} \ref{fig:bilstm_b}), the gate weights of $LSTM_{main}$ are flexible. At each timestamp, the auxiliary $LSTM_{aux}$ takes the psychophysiological parameters and spectrogram embedding to dynamically generate the $LSTM_{main}$’s gate weights for that corresponding timestamp. 
This means that the $LSTM_{main}$ network generates the class inferences, taking only fluent speech spectrograms as input; however, the $LSTM_{main}$’s gate weights that identify each timestamp’s speech information importance are adaptive and generated utilizing the speaker's physiological parameters. 
Such architecture enables applying SHAP on the $LSTM_{main}$ network and identifying the important speech attributes from the participant's fluent speech spectrogram (Discussed in Appendix \ref{appendix:shap}).
\par 
\textit{PASAD}'s overall architecture is a bi-directional \textit{HyperNetwork} structure composed of a forward \textit{Hyper-LSTM} and a backward \textit{Hyper-LSTM} (shown in \textit{Fig.} \ref{fig:bilstm_a}). The forward and backward \textit{Hyper-LSTM} unit output is concatenated and fed into a \textit{CWS} vs. \textit{CWNS} inference generator classifier composed of linear layers and a softmax function. The components of the \textit{Hyper-LSTM} unit are:
\par
\textbf{\textit{Hyper-LSTM} Unit: } 
In the presented structure at each timestamp $t$, feature-extractor and reference-extractor blocks take the physiological parameters ($P_t$) and speech spectrogram representations ($S_t$) as input. The $LSTM_{aux}$ takes their generated embeddings as input and generates the hidden state $\hat{h}_t$ that is used to generate the weights of $LSTM_{main}$ at the same timestamp. Both LSTMs are jointly trained with backpropagation and gradient descent.
\par
The $LSTM_{main}$ has four gates {$i, g, f, o$}. And each of the gates has unique weight matrices including $W_{h}$,$W_{x}$, and $b$. To generate weight matrices of $LSTM_{main}$, first, we use linear projection to map the hidden states $\hat{h}_t$ of the $LSTM_{aux}$ Cell to the embedding vectors unique to each gate: $z^{y}_{h}$,$z^{y}_{x}$, and $z^{y}_{b}$ where $y \in\{i, g, f, o\}$. We dynamically compute the weight matrices
of four gates as a linear transformation of $z$ as shown in equation \ref{eq:hyper}. 
\begin{equation}
\begin{aligned}
W^{y}_{h} = <W^{y}_{hz},z_{h}>, 
W^{y}_{x} = <W^{y}_{xz},z_{x}>, 
b^{y} = W^{y}_{bz}z^{y}_{b}+b^{y}_{0}
\end{aligned}
\label{eq:hyper}
\end{equation}
$<\cdot>$ is tensor product operation. $y \in\{i, g, f, o\}$, $W^{y}_{hz}$,$W^{y}_{xz}$ and $W^{y}_{bz}$ are parameters that can be learned during training. In this approach, the transformation weight tensor $W^{y}_{hz}$,$W^{y}_{xz}$ and $W^{y}_{bz}$ are pretty large. E.g.,  when $W^{y}_{h}$ has shape $N_{h}\times N_{h}$, The $W^{y}_{hz}$ will has shape $N_{h}\times N_{h}\times N_{z}$. To overcome this, we compute the weight parameters of $LSTM_{main}$ by dynamically scaling each row of a matrix of the same size as shown in equation \ref{eq:scale}.
\begin{equation}
W(z)=W(d(z))=\left(
\begin{aligned}
d_{0}(z)&W_{0} \\
d_{1}(z)&W_{1}\\
\dots \\
d_{N_{h}}(z)&W_{N_{h}} 
\end{aligned}
\right)
\label{eq:scale}
\end{equation}
where $W$ is $N_{h}\times N_{h}$ parameter matrix and $W_{i}$ is the $i$th row of matrix $W$. The $d(z)$ is a linear projection of $z$ and we refer $d$ as a weight scaling vector where $d_{i}$ is  $i$th element of the vector. With the weight matrices of four gates, we can perform LSTM operation on the main data flow as shown in equation \ref{eq:LSTM}.
\begin{equation}
\begin{aligned}
c_{t}=\sigma(f_{t}) c_{t-1} + \sigma(i_{t}) tanh(g_{t}), h_{t}=\sigma(o_{t}) tanh(LN(c_{t})) 
\end{aligned}
\label{eq:LSTM}
\end{equation}
where $c_{t}$ and $h_{t}$ are memory cell and hidden state, and $i_{t},f_{t},g_{t},o_{t}$ are outputs of the four gates of $LSTM_{main}$ at timestamp $t$. $LN$ is layer normalization, which encourages better gradient flow, which helps stabilize the hidden state.
 Notably, the weight-generating embedding vectors $z$ are not constant. Hence, the weight of the $LSTM_{main}$ is different at each timestamp for different inputs depending on the speech and physiological characteristics. 
\par 
\textbf{\textit{Feature-Extractor: }} 
At each timestamp $t$, the feature extractor takes a $500$ms Mel-spectrogram $x_t$ as input. We introduce a non-local block similar to \citet{wang2018non} in the feature-extractor to capture long-range spatial-temporal relations, followed by convolution layers to extract abstract features. Later, the output of the convolution layers is flattened and represented as a 1D spectrogram embedding at timestamp $t$. \textit{The non-local block} encodes the \textit{pairwise} relations of all possible positions (frequency and time) of the Mel-spectrogram. Additionally, it encodes the \textit{unary} information, meaning how a position may have its independent impact on all other positions \cite{yin2020disentangled}. The formal definition of non-local operation is:

\begin{equation}
y_{i}=\frac{1}{C(x)}\sum_{\forall j}f(x_{i},x_{j})g(x_{j})
\label{eq:non-local}
\end{equation}
Where $x$ and $y$ are the non-local block's input and output. The $C(x)$ is the normalization factor. When computing the output at position $i$ ($y_i$), we enumerate all possible position $j$ of $x$. $f(x_{i},x_{j})$ represent\textcolor{red}{s} the \textit{pairwise} relation between $x$'s position i and j. It can be distance or affinity. Here, we use the Embedded Gaussian function as function f:
\begin{equation}
f(x_{i},x_{j})=e^{\theta (x_{i})^{T} \phi (x_{j})}
\label{eq:non-local_f}
\end{equation}
Here, $\theta(x_{i}) =W_{\theta}x_{i}$ and $\phi (x_{j})=W_{\phi}x_{j}$ are embeddings of $x_i$ and $x_j$, $W_{\theta}$ and $W_{\phi}$ are learnable network parameters. In our implementation, $W_{\theta}$ and $W_{\phi}$ are single-convolution-layers with kernel size of $1\times1$.
\begin{figure}[h]
\centering
\includegraphics[width=0.5\textwidth]{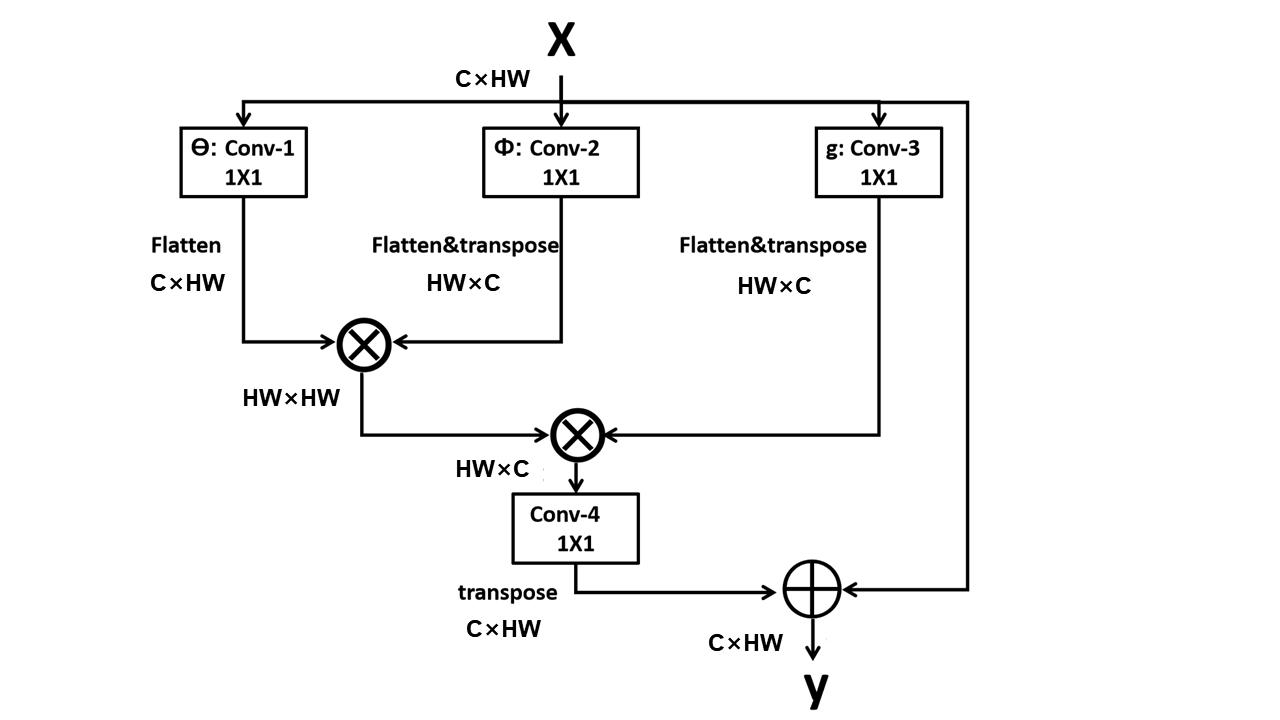}
\caption{The Nonlocal Block. $\otimes$ and $\oplus$ denote matrix multiplication and element-wise sum. ``Conv" denote $1\times1$ convolution layers. $C$ is the number of channel, $H$ and $W$ are height and width of input feature.}
\label{fig:non-local structure}
\end{figure}
\\$\theta (x_{i})^{T} \phi (x_{j})$ is the dot-product similarity. The normalization factor is set as $C(x)=\sum_{\forall j}f(x_{i},x_{j})$. With this equation, the normalization operation is equal to a softmax operation along the dimension j. 
\par 
The $g(x_j)$ represent a simple linear embedding of $x_j$:
$g(x_{j})=W_{g}x_{j}$, where $W_{g}$ is a learnable network parameter. In our implementation, $W_{g}$ is a single-convolution-layer with a kernel size of $1\times1$.
\par


Figure \ref{fig:non-local structure} illustrates the architecture of the non-local block. The input signal $x$ is passed through three $1 \times 1$ convolution layers, corresponding to the functions $\theta$, $\phi$, and $g$. The embeddings produced by these functions are flattened to one dimension. The output from $\phi$ (Conv-2) is transposed and then matrix-multiplied with the output from $\theta$ (Conv-1), performing the pairwise computation as described in equation \ref{eq:non-local_f}. A softmax operation follows to normalize the resulting scores. The output from $g$ (Conv-3) provides the linear embedding of the input, which is transposed and matrix-multiplied by the softmax output. This result captures the unary information, reflecting the independent relationship of each position with the combined correlations of all other positions. A final $1 \times 1$ convolution and a transpose operation are applied, resulting in an output $y$ with the same shape as the input $x$.

Additionally, the residual connection makes the module an independent block, allows gradient flow during back-propagation, and improves the training efficiency.
\par
\textbf{\textit{Reference-Extractor: }}
Reference-extractor block comprises a pair of non-linear layers that take the 1-dimensional change-scores as input and generates a 1D physiological feature embedding at each timestamp $t$.

\section{Classification Approach Evaluation}\label{evaluation}
This section discusses models' performance (Section \ref{Eval-performance}), ablation evaluation (Section \ref{ablation-eval}) of the \textit{PASAD}, and comparison of using raw physiological and acoustic features. Later, we empirically validate \textit{PASAD}’s dynamic weight learning through visualization in Section \ref{weight-visualization} and the absence of any information leakage to ensure \textit{PASAD}’s interpretation-visualizations are valid in Section \ref{leakage-eval}, age and sex-wise performance comparisons in Section \ref{eval-age-sex}, and execution and resource use on resource-constraint platforms in Section \ref{real-time-eval}. 
\par
Following the state-of-the-art works, we considered the CNN-LSTM \cite{hochreiter1997long}, Resnet-9 \cite{li2016demystifying}, Bi-directional LSTM networks \cite{chen2017improving}, and transformer \cite{vaswani2017attention} as baseline classifiers. PASAD performs classification considering only the speech modality, but it utilizes physiological parameters for better acoustic attribute analysis. Hence, for a fair comparison, for each baseline, we considered both \textit{multi-modal} and \textit{speech} classifiers taking \textit{spectrogram and change-score features} and the only \textit{spectrogram} as input.
The network implementations, experimental setup, system, and configurations are discussed in detail in Appendix \ref{appendix:implementation}.
\par 
\textbf{\textit{ Dataset Split and Evaluation Metrics:}} For each evaluation, we followed \textbf{\textit{the person-disjoint hold-out method}} \cite{cawley2010over}. Both \textit{stress-speech} and \textit{narrative-speech} datasets contain data from different children. We perform 10 fold-cross validation on both datasets. For each fold, We randomly separate each dataset into person-disjoint test ($2$ children: $1$ \textit{CWS}, $1$ \textit{CWNS}), validation ($8$ children: $4$ \textit{CWS}, $4$ \textit{CWNS}), and training (the rest of the children) subsets to avoid personal bias and make the models generalizable. The presented results are averaged over the ten groups to reduce contingency and avoid overfitting of the model. Evaluation results are presented with the metrics sensitivity, specificity, accuracy, and F1-score. 

\subsection{Classification Performance}\label{Eval-performance}

As shown in \textit{Table} \ref{table:detectiionperformance}, \textit{PASAD} achieves a relatively higher F1-score of $0.83$ in the \textit{stress-speech} dataset than the \textit{narrative-speech} dataset ($0.78$ F1-score). Children experienced higher arousal during the \textit{stress-speech} task (avg. EDA change-score was 4 times higher), which may result in lesser stable speech construction by the \textit{CWS} \cite{tumanova2020effects}, hence making the classification task easier in the \textit{stress-speech} dataset. Moreover, a sentence (i.e., BBAP phrase) was repeated in the \textit{stress-speech} dataset. A classifier could easily learn the acoustic characteristics of the BBAP phrase from \textit{CWS} vs. \textit{CWNS}, which was not the case in the \textit{narrative-speech} dataset. Hence, \textit{PASAD} performs relatively higher in the \textit{stress-speech} dataset.
\par
\textit{Tables} \ref{table:detectiionperformance} and \ref{evalspeech-baselines} show the evaluations of multi-modal and speech baselines. The best-baseline, Transformer, is a temporal-pattern-extracting sequential classifier \textit{that further justifies \textit{PASAD}’s sequential network structure.} 
Moreover, the speech baselines have relatively lower efficacy than the multi-modal baselines. It indicates that \textit{even experiencing similar situations, \textit{CWS} and \textit{CWNS} show distinctive physiological response patterns} that multi-modal classifiers can leverage to achieve higher performance. 
\begin{table}[]
    \parbox{.49\textwidth}{
        \centering
        \resizebox{0.45\textwidth}{!}{
        \begin{tabular}{|c|c|c|c|c|c|}
\hline
Dataset                                                                & Model              & Sensitivity & Specificity & Accuracy & F1 score \\ \hline
\multirow{5}{*}{\begin{tabular}[c]{@{}l@{}}\textit{Stress} \\ \textit{Speech}\end{tabular}}                                              & \textit{PASAD}&	\textcolor{purple}{0.8742}&	\textcolor{purple}{0.7526}&	\textcolor{purple}{81.73\%}&	\textcolor{purple}{0.8358}\\ \cline{2-6} 

&CNN-LSTM&	0.7515&	0.6236&	69.16\%&	0.7216\\ \cline{2-6} 
&Resnet-9&	0.7822&	0.6794&	73.4\%&	0.7578\\ \cline{2-6} 
&Bi-LSTM&	0.819&	0.6933&	76.01\%&	0.7841      \\ \cline{2-6}                 
&transformer&	0.8312&	0.7282&	78.3\%&	0.8029     \\ \hline                                                                  
\multirow{5}{*}{\begin{tabular}[c]{@{}l@{}}\textit{Narrative} \\ \textit{Speech}\end{tabular}} & \textit{PASAD}	&\textcolor{purple}{0.7683}&	\textcolor{purple}{0.837}	&\textcolor{purple}{80.46\%}	&\textcolor{purple}{0.7878}\\ \cline{2-6} 

&CNN-LSTM& 0.5812&	0.8087&	70.13\%&	0.6475\\ \cline{2-6} 
&Resnet-9&0.6722&	0.7723&	72.51\%	&0.6979\\ \cline{2-6} 
&Bi-LSTM&0.6683&	0.837	&75.36\%&	0.7223\\ \cline{2-6} 
&transformer&0.7316&	0.8193&	77.79\%&	0.7567\\ \hline
\end{tabular}}
        \caption{\textit{PASAD}'s classification performance compared with multi-modal baselines.  \label{table:detectiionperformance}}
    }
    \hfill
    \parbox{.49\textwidth}{
        \centering
        \resizebox{0.45\textwidth}{!}{
        \begin{tabular}{|c|c|c|c|c|c|}
\hline
Dataset                                                                & Model              & Sensitivity & Specificity & Accuracy & F1 score \\ \hline
\multirow{4}{*}{\begin{tabular}[c]{@{}l@{}}\textit{Stress} \\ \textit{Speech}\end{tabular}}                                              

&CNN-LSTM&0.7607&	0.6062&	68.84\%&	0.7219\\ \cline{2-6} 
&Resnet-9&0.8742&	0.3832&	64.43\%&	0.7233\\ \cline{2-6} 
&Bi-LSTM&0.7453&	0.5679&	66.23\%&	0.7012 \\  \cline{2-6} 
&transformer&0.8312&	0.6027&	72.43\%&	0.7623 \\ \hline

\multirow{4}{*}{\begin{tabular}[c]{@{}l@{}}\textit{Narrative} \\ \textit{Speech}\end{tabular}}

&CNN-LSTM&0.6613&	0.7909&	72.97\%&	0.698\\ \cline{2-6} 
&Resnet-9&0.6504&	0.8697&	76.62\%&	0.7243\\ \cline{2-6} 
&Bi-LSTM&0.6089&	0.8786&	75.12\%&	0.698\\  \cline{2-6} 
&transformer&0.6504&	0.8715&	76.71\%&	0.7251\\ \hline

\end{tabular}}
        \caption{Speech classification baselines' (only spectrogram as input) performance. \label{evalspeech-baselines}}}
          \vspace{-0.5cm}
\end{table}
\par
\textit{PASAD} utilizes a simple bi-directional LSTM architecture, and still utilizing only speech modality in decision making, it outperforms complex multi-modal models such as Transformer.
It is important to note that this paper’s classification objective is challenging since both \textit{CWS} and \textit{CWNS} are fluently speaking and experiencing similar situations; hence, the situational physiological response differences are subtle and not perceptually differentiable. The evaluation results demonstrate that \textit{PASAD} effectively extracts \textit{CWS} vs. \textit{CWNS} differentiating dynamic acoustic aspects by leveraging the physiological parameters. In contrast, the baseline multi-modal models comprise static network weights, and hence fail to extract the dynamic and subtle differentiating aspects of \textit{CWS} vs. \textit{CWNS} with similar efficacy, hence performing relatively poorly.
\par

\subsection{Ablation Study.}\label{ablation-eval} 
We performed an extensive ablation study of \textit{PASAD}'s architecture. Table \ref{table:AblationEVal} shows the ablation evaluation where different components are modified while keeping other components unchanged. We aim to classify using speech information; hence, the $LSTM_{main}$ takes only spectrogram as input, whereas the $LSTM_{aux}$ extracts physiological response aided temporal speech importance, taking both spectrogram and change-score as input. Without Non-local block reduces F1-scores $2-3$ percentage points. It indicates that the Non-local block effectively captures the long-range spatial-temporal relations of different frequencies and time, leading to higher performance.

 \textbf{\textit{Only Spectrogram or Change-score Embedding in the $LSTM_{aux}$:}} Existing studies \cite{ha2016hypernetworks,gui2018transferring} utilizing HyperNetwork structures put the same input in the main $LSTM_{main}$ and auxiliary $LSTM_{aux}$. We aim to classify using speech information; hence, the $LSTM_{main}$ takes only spectrogram as input, whereas the $LSTM_{aux}$ extracts physiological response aided temporal speech importance, taking both spectrogram and change-score as input. Without the physiological information (i.e., change-score) in $LSTM_{aux}$, PASAD achieves only $0.67$ and $0.76$ F1-scores in stress-speech and narrative-speech datasets, indicating the change-score feature’s efficacy in extracting CWS vs. CWNS differentiating temporal speech information. 
Using only change-score in the $LSTM_{aux}$, PASAD achieves $0.73$ F1-scores, indicating the need for $LSTM_{aux}$ to incorporate both acoustic and physiological features to generate the weight for $LSTM_{main}$. Possibly due to the mixed effect of acoustic and physiological features. This result demonstrates the effectiveness of the PASAD implementation.

\begin{table}

\centering
    \captionsetup[subtable]{labelformat=empty}
    \begin{minipage}{.9\linewidth}
   
        \resizebox{\columnwidth}{!}{%
        \begin{tabular}{|c|c|c|c|c|c|}
\hline
Dataset                                                                 & Modification                                                                                    & Sensitivity                                         & Specificity                                           & Accuracy                                              & F1 score                                              \\ \hline
                                                                        & PASAD                                                                                           & 0.8742&	0.7526&	81.73\%&	0.8358                                            \\ \cline{2-6} 
                                                                        & \begin{tabular}[c]{@{}c@{}}without\\ non-local\\ block\end{tabular}                             & 0.8435&	0.7317&	79.11\%&	0.8112                                                \\ \cline{2-6} 
                                                                        
                                                                        & \begin{tabular}[c]{@{}c@{}}With only \\ Spectrogram\\ embedding in\\ $LSTM_{aux}$\end{tabular}  & 0.7085&	0.5435&	63.13\%	&0.6715                                               \\ \cline{2-6} 
\multirow{-5}{*}{\begin{tabular}[c]{@{}c@{}}Stress\\Speech\end{tabular}}                                              & \begin{tabular}[c]{@{}c@{}}With only \\ Change-score\\ embedding in\\ $LSTM_{aux}$\end{tabular} & 0.7208& 	0.7317& 	72.59\%& 	0.7366 \\ \hline
                                                                        & PASAD                                                                                           &0.7683&	0.837	&80.46\%	&0.7878                                               \\ \cline{2-6} 
                                                                        & \begin{tabular}[c]{@{}c@{}}without\\ non-local\\ block\end{tabular}                             & 0.7237&	0.8724&	80.22\%	&0.7755                                          \\ \cline{2-6} 
                                                                    
                                                                        & \begin{tabular}[c]{@{}c@{}}With only \\ Spectrogram\\ embedding in\\ $LSTM_{aux}$\end{tabular}  & 0.7624& 	0.7839& 	77.37\%	& 0.7609                                        \\ \cline{2-6} 
\multirow{-4}{*}{\begin{tabular}[c]{@{}c@{}}Narrative \\ Speech\end{tabular}} & \begin{tabular}[c]{@{}c@{}}With only \\ Change-score\\ embedding in\\ $LSTM_{aux}$\end{tabular} & 0.6822&	0.8512&	77.14\%&	0.7381                \\ \hline
\end{tabular}
        }
        \centering
        \hfill
        \subcaption{(a)Result for  ablation evaluation of PASAD model structure\label{table:AblationEVal}}\hfill
        \resizebox{\columnwidth}{!}{%
        \begin{tabular}{|c|c|c|c|c|c|c|}
\hline
Dataset                                                                & \begin{tabular}[c]{@{}c@{}}Audio\\ Features\end{tabular} & \begin{tabular}[c]{@{}c@{}}Physio.\\ Features\end{tabular} & Sensitivity & Specificity & Accuracy & F1 score \\ \hline
\multirow{2}{*}{\begin{tabular}[c]{@{}c@{}}Stress \\ Speech\end{tabular}}                                                      & Raw                                                      & \begin{tabular}[c]{@{}c@{}}Change\\ Score\end{tabular}     & 0.75        & 0.596       & 67.31\%  & 0.696    \\ \cline{2-7} 
                                                                       & Spectrogram                                              & Raw                                                        & 0.6923       & 0.5962       & 64.42\%  & 0.6606    \\ \hline
\multirow{2}{*}{\begin{tabular}[c]{@{}c@{}}Narrative \\ Speech\end{tabular}} & Raw                                                      & \begin{tabular}[c]{@{}c@{}}Change\\ Score\end{tabular}     & 0.733       & 0.332       & 50.9\%   & 0.564    \\ \cline{2-7} 
                                                                       & Spectrogram                                              & Raw                                                        & 0.7254       & 0.5042       & 60.2\%   & 0.627    \\ \hline
\end{tabular}
        }
        \centering
        \hfill
        
        \subcaption{(b) Result for Raw audio \& physiological features evaluation\label{table:raw-feat}}\hfill
        
        \hfill
        
    \end{minipage}
\caption{Result for Binary Classification (a) Ablation evaluation of PASAD model structure, (b) Raw audio \& physiological features evaluation.}
\label{tab:CM_all_binary_withandwithoutLateFusion}
    \vspace{-0.6cm}
\end{table}

\subsection{Raw Physiological Features.} Table \ref{table:raw-feat}’s evaluation shows the performance while utilizing the raw physiological features as input in the Reference-Extractor. Raw acoustic features are limited by the bias due to different individuals’ variable baseline condition physiological parameters. Conforming to that, the classifiers perform poorly utilizing the raw features. 
\par
\subsection{Raw Acoustic Features.} Table \ref{table:raw-feat}’s evaluation shows the performance while utilizing the raw acoustic features as input. Here, 2D CNN was replaced with 1D due to the input dimension. Such a model cannot extract the pairwise and unary relations across frequency and time of speech acoustic; hence performs relatively poorly.

\subsection{$LSTM_{main}$ Gate weights Visualization}\label{weight-visualization}
\begin{figure}

\begin{subfigure}{.2\textwidth}
\centering
  \includegraphics[width=.99\linewidth]{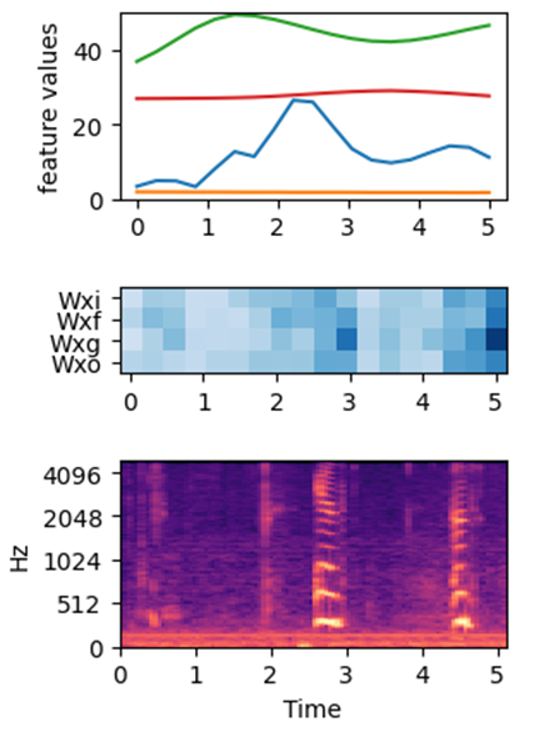}
  \caption{\textit{Narrative-speech}}
\end{subfigure}%
\begin{subfigure}{.3\textwidth}
\centering
  \includegraphics[width=.90\linewidth]{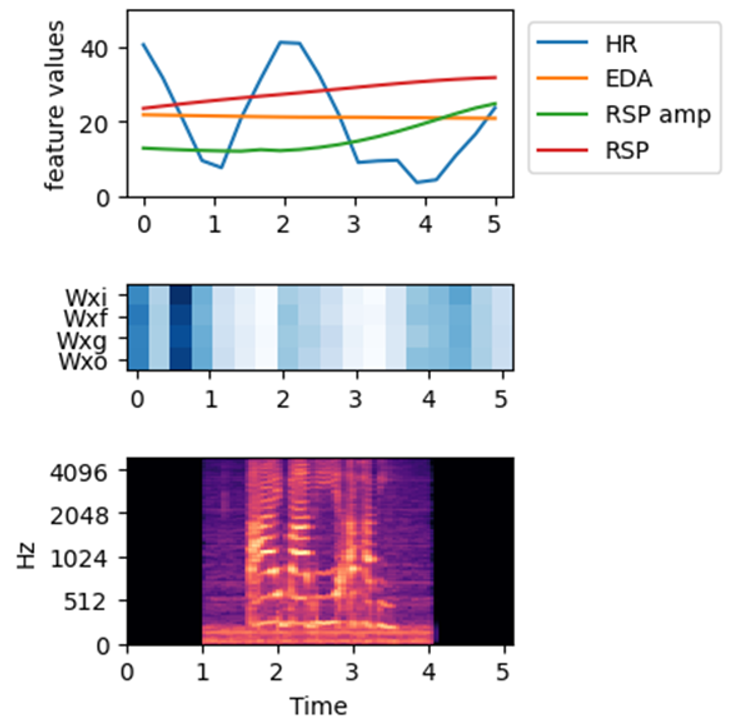}
  \caption{\textit{Stress-speech}}
\end{subfigure}
\caption{$LSTM_{main}$ Gate weight visualization with respective change-score features and Mel-spectrogram. The time metric is on seconds, and the darkness of the blue color shows gate weight change intensity.}
\label{fig:gate_fsall}
\vskip -3ex
\end{figure}

This section demonstrates the dynamic changes of the $LSTM_{main}$ gate weights (generated by $LSTM_{aux}$) with the changes in physiological and speech information.
\textit{Fig.} \ref{fig:gate_fsall} visualize the Mel-spectrogram and change-score features alongside the $LSTM_{main}$ weights changes for two examples. Euclidean distance between the $LSTM_{main}$ weight matrix of the previous and current timestamp is used to compute the gate weights changes. 
\par
In the \textit{narrative-speech} example, HR fluctuates in $1.5$-$3$ \&  $4$-$4.7$ sec; hence the gate weights have greater changes. In the \textit{stress-speech} example, participant’s utterance comprises 3 sec; hence $0$-$1$ \&  $4$-$5$ sec in the spectrogram are zero-padded. Gate weights change significantly in $1$ \& $4$ sec, where the zero-pad and speech transition occurs.
During $2$-$3$ \&  $4$-$5$ sec, the HR is rapidly changing, which causes greater changes in the gate weights. The discussion demonstrates that $LSTM_{main}$ gate weights are expressive and adaptive to the speech and change-score information changes.
\subsection{Data Leakage through Physiological Parameters}\label{leakage-eval}
Though the weights of $LSTM_{main}$ are adaptive to the speaker’s physiological responses, $LSTM_{main}$ performs classification using only the extracted Mel-spectrogram embeddings. Information leakage through physiological parameters is a concern, meaning if the $LSTM_{main}$ learns \textit{CWS} vs. \textit{CWNS} differentiating patterns independently from the change-score features (through the generated weights).
\par
To investigate, we train and evaluate the \textit{PASAD} approach while replacing the Mel-spectrograms with random noise (on similar frequency ranges) and keeping the actual change-score features. Such classifiers must learn \textit{CWS} vs. \textit{CWNS} differentiating patterns independently from the change-score feature. The evaluation is shown in \textit{table} \ref{table:randomnoise}, where the classifiers perform significantly poorly (i.e., random inferences), demonstrating the absence of information/data leakage.

\begin{table}[h]
\resizebox{0.45\textwidth}{!}{

\begin{tabular}{|c|l|l|l|l|}
\hline
Dataset     & \multicolumn{1}{c|}{Sensitivity} & \multicolumn{1}{c|}{Specificity} & \multicolumn{1}{c|}{Accuracy} & \multicolumn{1}{c|}{F1 score} \\ \hline
        \textit{Stress-Speech}    & 0.6346                           & 0.5                             & 56.73\%                        & 0.5946                        \\ \hline
\textit{Narrative-Speech} & 0.3208                           & 0.5212                           & 43.26\%                        & 0.3348                        \\ \hline
\end{tabular}

}
\caption{Evaluation of \textit{PASAD} by replacing Mel-spectrograms with random noise}
\label{table:randomnoise}
\vskip -4ex
\end{table}

\subsection{Age and Sex -Wise Analyze}\label{eval-age-sex}
This section discusses PASAD’s efficacy in different sex and age groups. For a fair comparison, we perform this section’s evaluation on the stress-speech dataset where all participants uttered the same `Buy Bobby a puppy,’ i.e., BBAP phrase experiencing the same condition. 
\par
Notably, in the stress-speech dataset, 9 children (2 female, 7 male) were 2-4 years of age, 16 children (3 female, 13 male) were 4-5 years of age, and 13 (all male) were over 5 years of age. 
The evaluation results are shown in Table \ref{age-wise scripted}. The relatively poor performance in the female group is due to the imbalanced male vs. female ratio in the dataset. However, PASAD achieves higher performance in the higher age group. Previous literature \cite{green2000physiologic} showed that speech-motor-controls are unstable in smaller age groups irrespective of whether they stutter or not. Hence, differentiating fluent speech of CWS vs. CWNS in the small age groups is difficult. With higher age, the speech-motor-control gets more stable; hence it is easier to identify the CWS indicative speech-motor-control instabilities in the higher age groups. Hence, PASAD’s performance is in line with the literature.
\par 
Importantly, this section’s evaluation demonstrates that PASAD generalizes and performs relatively uniformly in each age-sex group.

\begin{table}[]
\resizebox{0.45\textwidth}{!}{
\begin{tabular}{|l|l|l|l|l|}
\hline
  Sex & 3-4 years old & 4-5 years old  & over 5 years old & Average\\ \hline
  Female& 100\%&76.49\%&0&88.24\%\\ \hline
  Male&80.39\%&90.68\%&93.09\%&89.42\%\\ \hline
  Average&85.29\%&88.5\%&93.09\%&\\ \hline
 
\end{tabular}
}
\caption{Average accuracy sex-wise and age-wise in stress-speech dataset}
\label{age-wise scripted}
  \vskip -4ex
\end{table}

\subsection{Execution Time and Resource Usage }\label{real-time-eval}
Real-time executability is crucial for the CWSs' perceptually fluent speech anomaly assessment classifier, as it ensures the detection process aligns with the natural flow of conversation, allowing for timely intervention and feedback. In practical settings like speech therapy, classrooms, or assistive communication devices, rapid detection enables immediate support or adaptation, providing CWS with real-time assistance in managing their speech. To evaluate the PASAD's real-time executability and resource usage, we deploy the approach on an Nvidia Jetson Nano \cite{jetsonnano} and smart phone\cite{pixel6}. The jetson nano  equipped with NVIDIA Maxwell GPU, Quad-core ARM processor, and 4GB memory. The smart phone is Google Pixel 6 with  8gb ram and eight cores CPU. We run the models taking consecutive $5$ sec windows for $10$ minutes ($120$ repetitions), and record the running time and resource usage. As shown in table \ref{Running time}, both models take about 1 sec to process a $5$ sec speech information on both jetson nano and smartphone. The average CPU and GPU usages are $15$-$16$\% and $8$-$12$\% on jetson nano. And the average CPU usage on smart phone is $3$2\%-$33$\%. On the jetson nano, the average RAM usage is around 2.4Gb. For the smartphone execution, we optimize the model for mobile execution and the RAM usage is as low as around 0.33Gb; hence, the models do not significantly occupy the resource. The results suggest that PASAD can perform real-time analysis on resource constraint devices.

\begin{table*}[]
\resizebox{0.7\textwidth}{!}{
\begin{tabular}{|l|l|l|l|l|}
\hline
  Model &  \begin{tabular}[c]{@{}c@{}}stress-speech model \\ (jetson nano)\end{tabular}& 
  \begin{tabular}[c]{@{}c@{}}narrative-speech model \\ (jetson nano)\end{tabular}&
  \begin{tabular}[c]{@{}c@{}}stress-speech model \\ (smartphone)\end{tabular}&
  \begin{tabular}[c]{@{}c@{}}narrative-speech model \\ (smartphone)\end{tabular} \\ \hline
  Running time & 0.8966 s  & 1.0237 s  &0.580 s & 1.67 s  \\ \hline
  Average CPU usage & 15.06\%   &15.87\%  & 32\%&33\%\\ \hline
  
  Average GPU usage & 8.59\%   & 11.87\%  &N/A &N/A\\ \hline
  Average RAM usage &  2.4Gb/4Gb  & 2.5Gb/4Gb  &0.33Gb/8Gb &0.35Gb/8Gb\\ \hline

\end{tabular}
}
\caption{Average running time and resource usage}
\label{Running time}
  \vskip -6ex
\end{table*}
\section{\textit{PASAD} Model Interpretation Discussion} \label{Model interpretation}
This section discusses and demonstrates how \textit{PASAD}’s inferences can be utilized to understand stuttering children’s speech and speech-motor-control differences from the others. Understanding which Mel-spectrogram frequencies across time are important for each of the \textit{PASAD}’s inference generation is critical since they contribute most to differentiating a \textit{CWS}’s respective speech from others. We employ Kernel SHAP, a model-agnostic interpretation framework, that leverages \textit{PASAD}'s \textit{HyperNetwork} architecture to identify the critical Mel-spectrogram frequencies that $LSTM_{main}$ utilizes for \textit{CWS} vs. \textit{CWNS} inference generation.  
\par 
Notably, we are interested in generating the importance of different Mel-spectrogram frequencies (i.e., attributes). Hence, we divide the Mel-spectrogram into 32 frequency bands and use these bands as tokens for the Kernel SHAP. While generating the perturbation of the original Mel-spectrogram, we permute the tokens, meaning we replace all pixels within the frequency band with the background/other data and feed it into the kernel SHAP. After the Kernel SHAP analysis, each frequency band was assigned a Shapley (SHAP) value to represent its contribution to prediction. Finally, we visualize them with a horizontal bar plot. A detailed discussion of the Kernel SHAP employment procedure is in Appendix \ref{appendix:shap}. The following sections discuss how different Mel-spectrogram frequencies (e.g., formants, $F0$) refer to the deviation in different speech-motor-control components and demonstrate important distinct information identification from \textit{CWS}’s fluent speech through the \textit{PASAD}’s inference interpretation visualization.

\par 
\textbf{Mel-spectrogram Frequency Attributes and Speech-motor-control:}
Formants change with changes in the vocal tract configuration as speakers move their articulators (jaw, tongue) to produce a specific sound. Existing literature \cite{vorperian2019corner} evaluated the lowest four formants ($F1$-$F4$) in distinguishing unique vowel sounds. $F1$ \& $F2$ are associated with the position of the tongue. The height of the tongue in the mouth is inversely related to $F1$, such that high vowels have lower $F1$ values and vice versa. The $F2$ is directly related to tongue advancement (how far back the tongue is in the mouth). Back vowels have lower $F2$ values, and vice versa. 
$F3$ \cite{lindblom1979formant} is associated with the front cavity formed by the lips and the tongue constriction, and $F4$ \cite{vorperian2019corner} is associated with laryngeal descent/elevation and the pharyngeal and hypopharyngeal cavities. Additionally, lower speech-motor-control is positively correlated with higher fundamental frequency $F0$ variability \cite{robb1985developmental}. 
Also, deviation in high frequency indicates deviation to aerodynamic and resonance properties in the vocal tract (i.e., tongue placement) \cite{denny2000respiratory}.

\begin{figure*}
\centering
\begin{subfigure}{0.7\textwidth}
  \centering
  \includegraphics[width=.99\linewidth]{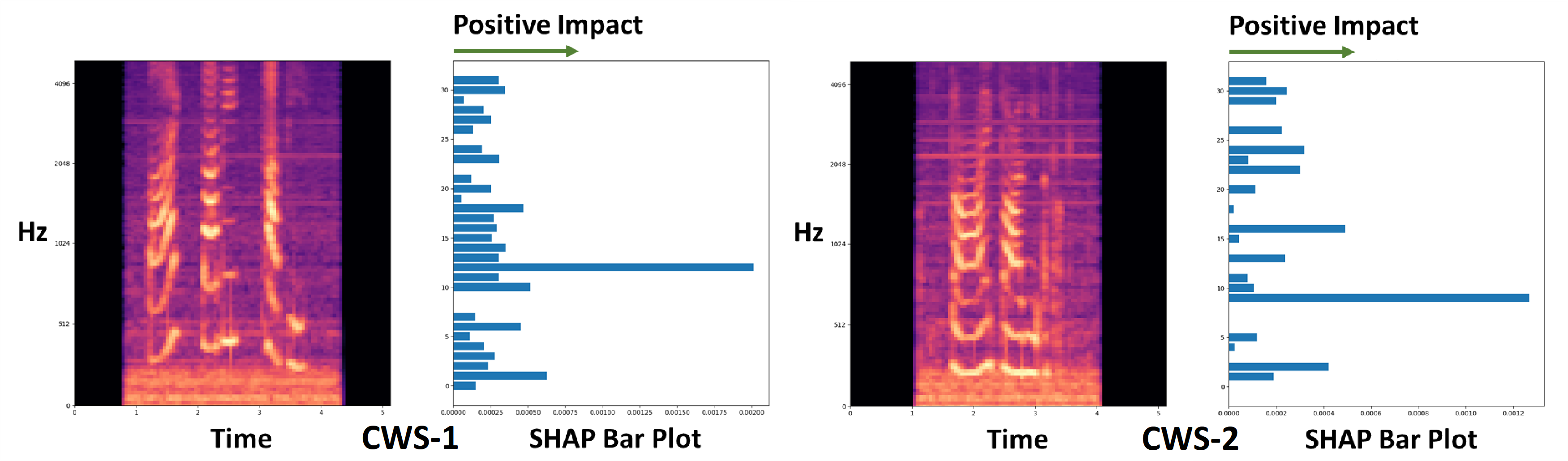}
  \caption{\textit{CWS} samples from \textit{stress-speech} dataset}
    \label{fig:SHAP-scripted}
\end{subfigure}
\begin{subfigure}{0.7\textwidth}
  \centering
  \includegraphics[width=.99\linewidth]{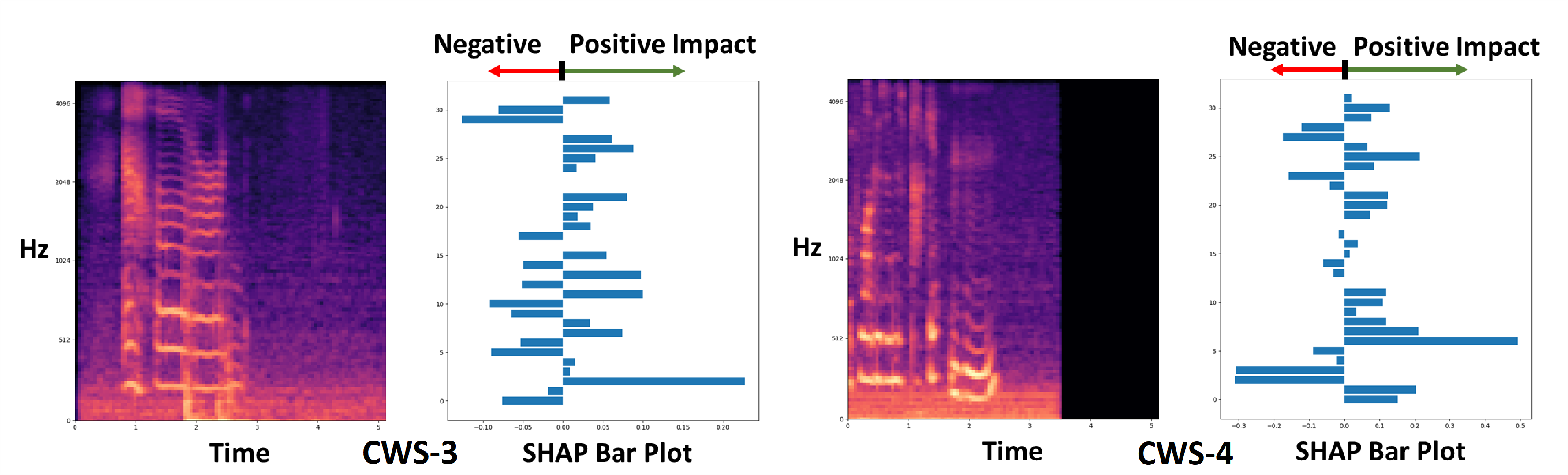}
  \caption{\textit{CWS} samples from \textit{narrative-speech} dataset}
  \label{fig:SHAP-FS}
\end{subfigure}
\vskip -1ex
\caption{Visualization of important spectrogram coordinates for \textit{CWS} speech samples}
\label{fig:SHAPinterpret}
\vskip -3ex
\end{figure*}
\subsection{Interpretation Discussion}\label{Interpretation:discussion} 
\textit{Fig.} \ref{fig:SHAP-scripted} and \ref{fig:SHAP-FS} show the \textit{PASAD} model Shapley interpretation of different \textit{CWS} participants' speech. In each figure, the left-side is the Mel-spectrogram, and the right-side is the Shapley importance bar-plot for different frequencies. 
\par
According to \textit{Fig.} \ref{fig:SHAP-scripted}, both \textit{CWS} participants’ BBAP phrase speech have higher $F0$ (0-300Hz) variability indicating lower speech-motor-control and higher physiological arousal. It is in line since they are experiencing higher arousal during this stress-speaking task.
Moreover, the Shapley importance is higher for $F3$, indicating deviation in their cavity formation by the lip and tongue constriction. 
\par 
According to \textit{Fig.} \ref{fig:SHAP-FS}, the $F1$ has higher Shapley importance for both \textit{CWS} participants from the \textit{narrative-speech} dataset. It indicates that they have deviations in tongue positions while speaking vowels. Participant \textit{CWS}-3 has higher Shapley values associated with $F4$, indicating deviation associated with laryngeal descent/elevation and the pharyngeal and hypopharyngeal cavities. Participants \textit{CWS}-4 has higher Shapley values associated with $F3$, indicating deviations in cavity formation by lip and tongue constriction.
Both participants have relatively higher Shapley values in higher frequencies, indicating minor deviation in tongue placement.
\par
Finally, this section demonstrates that the presented approach can identify important speech-motor-control information from \textit{CWS}'s fluent speech, making it impactful for practical use.  The ability to analyze personalized and subtle variations in speech patterns allows speech-language pathologists to identify key acoustic features associated with speech-motor control anomalies in \textit{CWS}, such as reduced fluency, muscle tension, or disrupted speech rhythm. By understanding how a child's speech deviates from typical development, \textit{PASAD} can support the creation of personalized, data-driven treatment plans tailored to each \textit{CWS}’s needs.

\section{Discussion on Broader Impact and Limitations}\label{discussion}
\textbf{Dataset Task Conditions: } Speech science studies \cite{byrd2012speech} use structured tasks that can elicit specific behavior-related data from the stuttering individuals in which clinicians are interested. Particularly, speaking in stress conditions \cite{guttormsen2015communication} and narration tasks \cite{zackheim2003childhood} are of interests due to the elicitation of arousal due to external stressors and cognitive loads. Following the existing studies, \textit{PASAD} identifies and visualizes the fluent speech attributes that distinguish \textit{CWS} vs. \textit{CWNS} in these two conditions. 
\par 
\textbf{Practical Impact:} This study's collected sensing modalities are present in recent wearables (e.g., \citet{biopacmobile}). According to our evaluation, \textit{PASAD} is real-time executable in smart devices. Hence, \textit{PASAD} has the potential to be implemented in wearables for remote and continuous assessment of \textit{CWS}’s fluent-speech and speech-motor-control deviation, and may lead to personalized and just-in-time interventions.
\par
\textbf{Limitation:} The study's limitation is that we analyzed data from only two conditions. Future work would benefit from sampling data from a wider range of situations to determine the boundaries of the models’ predictive validity. Additionally, \textit{PASAD} can be evaluated on wearables for the longitudinal understanding of \textit{CWS}’s speech characteristics. 

\section{Conclusion}\label{Conclusion}
To our knowledge, \textit{PASAD} is the first approach that leveraged the \textit{HyperNetwork} structure on multi-modal human sensing data. \textit{PASAD} utilizes the speaker’s physiological response for effective acoustic analysis and makes classification decisions using only speech information; ergo, it is not a multi-modal sensor data integration approach but a speech analysis approach. However, for a fair comparison, this paper compares \textit{PASAD} with multiple multi-modal classifiers and speech classifiers, all of which \textit{PASAD} outperforms. 
Its high efficacy establishes that \textit{CWS}’s distinct speech-motor-control is dynamic in nature and conforms to the respective speaker’s situational emotional reactivity. 
Our comprehensive evaluation of multiple speech conditions, ablation evaluation, and several baselines establish \textit{PASAD}’s efficacy, expressiveness, and generalizability. 
Furthermore, the presented interpretation discussion demonstrates \textit{PASAD}'s potential impact in enabling personalized speech-motor-control related intervention generation to \textit{CWS}.

\bibliographystyle{ACM-Reference-Format}
\bibliography{sample-base}

\newpage

\appendix

\section{Physiological and Acoustic Features}
\label{appendix:feature}
\subsection{Physiological Change-score Features:} State-of-the-art behavioral science studies evaluate change-score \cite{chiu2015complexity,clifton2019correlation,jones2017executive,sharma2022a} features to understand the psychophysiological changes on individuals in response to different affective states (e.g., stress, arousal, etc.). A change score is a difference between the value of a variable/feature measured at one point in time ($Y_t$) from the average value of the variable for the same unit at the baseline-level condition ($Y_b$). $Y_t$ is called the `post-score,’ $Y_b$ is the `baseline-score,’ and the difference between $Y_t$ \& $Y_b$ are the `change-scores’. This study extracts change-scores of HR, EDA, RSP-amp, and RSP-rate LLD features from each $500$ ms segment.
The post-, baseline-, and change-scores of these physiological features are represented as vectors.
\par 
\textbf{\textit{Post-scores}} are calculated from each $500$ ms physiological-signal-segments in different non-baseline scenarios. During raw features extraction, each of the LLD features is represented as a 6-dimensional vector (i.e., one dimension for each HLD-functional) quantifying an individual's physiological response under the respective $500$ ms time-segment. These 6-dimensional vectors $\vec{Y_{HR}}$, $\vec{Y_{EDA}}$,  $\vec{Y_{RSP-amp}}$, $\vec{Y_{RSP-rate}}$ are the respective LLD features' post-scores.
\par 
\textbf{\textit{Baseline-scores}} are calculated from all of the $500$ ms physiological-signal segments in the individual's baseline condition. For each LLD feature (HR, EDA, RSP-amp, and RSP-rate), we consider the mean of its 6-dimensional HLD vectors from all baseline-condition $500$ ms segments as its `baseline-score' vector. Meaning, we extract four 6-dimensional baseline-scores (i.e., $\vec{B_{HR}},\vec{B_{EDA}},\vec{B_{RSP-amp}},\vec{B_{RSP-rate}}$) for each individual, representing the average LLD-features values in their baseline (i.e., neutral) condition.
\par 
\textbf{\textit{Change-scores}} are the vector differences between the post-scores and baseline-score. For each $500$ ms non-baseline segment, it quantifies the difference in an individual's physiological response regarding their baseline condition (i.e., neutral) response. 
\textit{Different individuals may have different physiological responses in the baseline condition. For example, an individual may have above-average EDA in the baseline condition. However, a classifier trained on raw physiological features would be unaware of such exceptions and infer the individual's neutral state as a stress state. The change-score features eliminate such biases.}
\par 
In this study, the vector difference between the post-score and baseline-score is measured by two matrices: cosine similarity that measures the cosine of the angle between two vectors and the euclidean distance \cite{kryszkiewicz2014cosine,gomez2019self,wang2005euclidean}. For each of the four LLD features HR, EDA, RSP-amp, RSP-rate, we calculate two distance measures, totaling eight change-score features extracted from each $500$ ms small-signal segment. In total, we extract $19\times8$ change-score features from the $5$ sec window.

\subsection{Acoustic Mel-Spectrogram Features:} Spectrogram represents the speech's strength (i.e., energy) over time at various frequencies. We extract a Mel-spectrogram from each $500$ ms audio segment. It represents the acoustic signal on the Mel scale. The Mel Scale is a logarithmic transformation of a signal’s frequency, such that sounds within equal distance on the Mel Scale are perceived as equal to humans. 
State-of-the-art acoustic analysis studies \cite{suhas2020speech,bhattacharya2021animal,dieleman2014end} suggested that the use of Mel-spectrogram leads to improved accuracy than the conventional spectrograms.
During the time to frequency domain transformation, the length of the FFT window is 2048 sample points and hop length is 512 sample points. The desired frequency resolution governs the choice of FFT bins. Short-term spectral measurements are carried out for smaller window sizes of $20$ ms \cite{deller2000discrete,benesty2008introduction}. This controls the trade-off between the time and frequency resolution for the signal. At each timestamp, a Mel-spectrogram from a $500$ ms speech signal is input to the detection network's `Feature-Extractor' block of PASAD.

\subsection{Acoustic Raw Features:} This paper also evaluates the efficacy of using raw acoustic features. The first 13- Mel-frequency cepstral coefficient (MFCC), Zero crossing rate, fundamental frequency, and first four formant frequencies, and their delta coefficients extracted from each $500$ ms speech segment as the the low-level descriptor (LLD) features. Next, the 6 high-level descriptors (HLD) functionals: min, max, std, var, mean, median are applied on the LLDs to extract the feature representation of the $500$ ms segment.


\section{PASAD and Baseline Network Implementations}
\label{appendix:implementation}
\textbf{\textit{PASAD Implementation:}} PASAD approach has four components discussed below. The specific number of layers and number of neurons was set as hyper-parameter and optimized by a python toolkit "Optuna" \cite{akiba2019optuna}. It uses a Bayesian Optimization algorithm called Tree-Structured Parzen estimator to identify the optimum set of values. 
\begin{itemize}
\item \textit{The Hyper-LSTM unit} comprises of a one-layer auxiliary $LSTM_{aux}$ that generates a 1-dimensional hidden state $\hat{h}_t$. We use linear projection (i.e., one-linear layer) to map $\hat{h}_t$ to the embedding vectors unique to each gate: $z^{y}_{h}$,$z^{y}_{x}$ and $z^{y}_{b}$ where $y \in\{i, g, f, o\}$. 
Each of the four gate weights is generated with $z$ and a trainable parameter $W$ following equation $2$ in the main paper. During optimization, we set the number of neurons in $LSTM_{Aux}$ and $LSTM_{main}$ range from 200 to 800 and 600 to 1200 separately. And the dimension of the weight-generating embedding vector ranges from 200 to 800. 
\item \textit{The Feature-Extractor network} comprises $1$ to $4$ non-local block and each followed by a convolution layer. There are also $5-12$ convolution layers and 2 linear layers after those non-local blocks with ReLU activation. Batch normalization are used after each convolution layer. During training, we set the dimension of embedding range from 200 to 800, and the channel of convolutional layers ranges from 8 to 128.
\item \textit{The Reference-Extractor} are two linear layers with ReLU activation for the stress-speech and narrative-speech dataset implementations. The dimensions of linear layers range from 200 to 800.
\item The output of forward and backward Hyper-LSTM are fed into an inference generator \textit{classification network}. The classification networks comprise three linear layers with dropout and ReLU activations for both the stress-speech and narrative-speech dataset implementations. The number of neurons in linear layers range from 600 to 1200. 
\end{itemize}

\textbf{\textit{Baseline Implementations:}} Following the state-of-the-art works, we considered the CNN-LSTM\cite{hochreiter1997long}, Resnet-9 \cite{li2016demystifying}, Bi-directional LSTM networks\cite{chen2017improving}, and transformer\cite{vaswani2017attention} as baseline classifiers. 
For each network, we considered both multi-modal and speech classifiers taking spectrogram and change-score features, and only spectrogram as input.
\par
In the multi-modal evaluations, the CNN-LSTM model comprises a spectrogram embedding extractor block consisting of $3$ to $5$ convolution layer followed by $2$ linear layers, a change-score embedding extractor block comprises of two linear layers, a single-layered LSTM, and a classifier comprises of two linear layers. The spectrogram and physiological features go through spectrogram and change-score embedding extractor block respectively. Then their embedding was concatenated and feeds into a single-layer LSTM. The classifier takes the last hidden state of LSTM as input and makes a classification. All linear and convolution layers used the ReLU activation function. Each convolution layer is followed by batch normalization. The Resnet-9, bidirectional LSTM, and transformer share a similar structure as the CNN-LSTM. The ResNet-9 replaces the convolution layers in the spectrogram embedding extractor with residual network layers. The bidirectional LSTM and transformer replace single-layered LSTM with bidirectional LSTM and transformer encoder respectively. 

In the speech classifier evaluations, the baseline models take only spectrogram as input. The spectrogram feature goes through the spectrogram embedding extractor block, sequential model, and classifier. The embedding extractor block, sequential model, and classifier followed the same configuration as multi-modal does.
\par
\textbf{System Configuration} All evaluations are carried in a distributed computing system, comprising 14,000 cores and 20 terabytes of memory. The system is optimized to perform large number of smaller parallel jobs. We submit the program as a job and each job will be assigned 2 CPUs and a GPU. The GPU was randomly selected from a resource pool including NVIDIA RTX A6000, NVIDIA RTX 6000, NVIDIA RTX 5000, NVIDIA GeForce GTX 1080 Ti, and NVIDIA GeForce GTX 750 Ti. 

\textbf{Training Configuration} We perform 10 fold-cross validation on both stress-speech and narrative-speech datasets And run Optuna for each test subset independently. To ensure reproducibility, we set the random seed as 2021 for the stress-speech task and 2022 for the narrative-speech task. We also set the pytorch using deterministic algorithms to limit the number of sources of nondeterministic behavior. During training, we try the learning rate range from 1e-6 to 1e-4. The drop-out rate of linear layers ranges from 0.1 to 0.3. The batch size range from 5 to 10. 

\textbf{Finalized hyper-parameters} Here we present examples of finalized hyper-parameters for stress-speech task and narrative-speech task. In the stress-speech dataset, we achieve the best performance with the following parameters: the dimension of spectrogram embedding is 533, the number of the channels of convolutional layers is 62, the number of convolutional layers is 7, the number of the non-local block is 1, the dimension of reference embedding is 506, the number of neurons of $LSTM_{aux}$ and $LSTM_{main}$ is 234 and 915, the dimension of weight-generating embedding vector is 487. The learning rate is 6.979e-05, the drop-out rate is 0.2915, and the batch size is 10. In the narrative-speech dataset, we achieve the best performance with the following parameters: the dimension of spectrogram embedding is 631, the number of the channels of convolutional layers is 105, the number of convolutional layers is 6, the number of the non-local block is 2, the dimension of reference embedding is 249, the number of neurons of $LSTM_{aux}$ and $LSTM_{main}$ is 378 and 1078, the dimension of weight-generating embedding vector is 715. The learning rate is 2.078e-06, the drop-out rate is 0.2125, and the batch size is 5.
\par
\textbf{Important Note:} The codes, extracted features, and data samples are submitted for reproducibility. 

\section{Kernel SHAP Interpretation Generation} 
\label{appendix:shap}
Kernel SHAP approximates the conditional expectations of Shapley values in deep learning models by perturbing the input (i.e., different portions/attributes of input) and observing the impact on the output \cite{lundberg2017unified}. Shapley values \cite{shapley1953stochastic} correspond to the contribution of each input portion towards pushing the model inference closer or away from the true class value. 
\par
The kernel SHAP trains a surrogate weighted linear regression model based on artificial samples generated by perturbing the input (of which we are generating interpretation) and the original deep learning model's inferences on those perturbed samples \cite{rodriguez2020interpretation}. Then use the coefficients of the linear regression model to determine the input feature (i.e., attributes) importance.
However, we cannot directly employ the Kernal SHAP in our PASAD model. PASAD takes both Mel-spectrogram and change-score physiological information as input, and kernel SHAP’s linear regression model trained on such combined input would not be able to differentiate the independent impact of the Mel-spectrogram from change-score information. Notably, we are only interested in interpreting the Mel-spectrogram information’s impact on PASAD’s inferences (that $LSTM_{main}$ uses for inference generation). The change score psychophysiology data is only for reference (generating weights of $LSTM_{main}$), meaning we need to exclude the auxiliary network $LSTM_{aux}$ when applying the kernel SHAP. 
\par 
The unique structure of PASAD enables us to address the challenge and extract the independent importance of Mel-spectrogram information in PASAD’s inference generation. In the presented PASAD model, the change-score psychophysiology data is only for generating the weight of $LSTM_{main}$. The $LSTM_{aux}$ takes the spectrogram and psychophysiology embeddings to generate the gate weights of $LSTM_{main}$. The $LSTM_{main}$ takes only the spectrogram's embedding and makes the inference.
Therefore, as the first stage of interpretation generation, we run the complete PASAD model with unperturbed original Mel-spectrogram (of which we are generating interpretation) and change-score input and save the gate weights of $LSTM_{main}$.
Later, in the second stage, Kernel SHAP is applied to the feature-extractor and $LSTM_{main}$ (with already learned gate-weights); where Kernel SHAP retrieves inferences of perturbed samples (of the original Mel-spectrogram), train linear regression model based on that, and generate the importance of different attributes of the original Mel-spectrogram.


\end{document}